\documentclass{ws-mpla}

\begin{document}

\markboth{C. Barbieri et al.}{One- and Two-Nucleon Structure}

\catchline{}{}{}{}{}

\title{One- and Two-Nucleon Structure form Green's Function Theory}

\author{\footnotesize C. Barbieri}
\address{Theoretical Nuclear Physics Laboratory, RIKEN Nishina Center, \\
               2-1 Hirosawa, Wako, Saitama 351-0198 Japan
}

\author{M. Hjorth-Jensen}
\address{Department of Physics and Center of Mathematics for Applications, University of Oslo, N-0316 Oslo, Norway}

\author{C. Giusti and F.D. Pacati}
\address{Dipartimento di Fisica Nucleare e Teorica dell'Universit\`{a} degli Studi di Pavia and Istituto Nazionale di Fisica Nucleare, Sezione di Pavia, I-27100 Pavia}

\maketitle


\begin{abstract}
We review some applications of self-consistent Green's function theory
to studies of one- and two-nucleon structure in finite nuclei.
 Large-scale microscopic calculations that employ realistic
nuclear forces are now possible.
Effects of long-range correlations are seen to play a dominant role
in determining the quenching of absolute spectroscopic factors.
They also enhance considerably ($e$,$e'$pn) cross sections in
superparallel kinematics, in agreement with observations.

\end{abstract}


\vspace*{4mm}

Single-particle (SP) states at the Fermi surface of shell closures [or
{\em quasiparticles} (QPs)] play a crucial role in nuclear structure.
Both SP energies and the interactions between states of two QPs are
essential inputs to standard shell-model (SM) calculations.
The strengths of QPs [the {\em spectroscopic factors} (SFs)] and the
many-body mechanisms that determine them have also direct implications
for the effective charges in the SM.
It follows that the evolution of SP properties with changing
proton-neutron asymmetry is central to the physics of exotic beams.
The behavior of SP energies has been linked to the 
averaged monopoles of the nuclear interaction.\cite{Ots.05,Ots.10}
Microscopic calculations of other QP properties will hopefully aid
in gaining greater insight into the structure of exotic nuclei.

This talk considers microscopic calculations of QPs properties
using Green's function theory in the Faddeev random phase approximation
(FRPA) method.\cite{bar.07,bar.09c}
 The FRPA is an expansion of the many-body problem in terms of
particle-vibration couplings which allows {\em ab-initio} calculations
based on modern realistic nuclear forces.
A similar formalism that could accommodate multiple vibrations has also
been considered recently in the form of Parquet theory.\cite{Ber.10}
Here, we discuss two recent applications of FRPA to the
one- and two-body spectral functions.

\section*{Single-particle spectral function}

\begin{figure}[t]
\centerline{\psfig{file=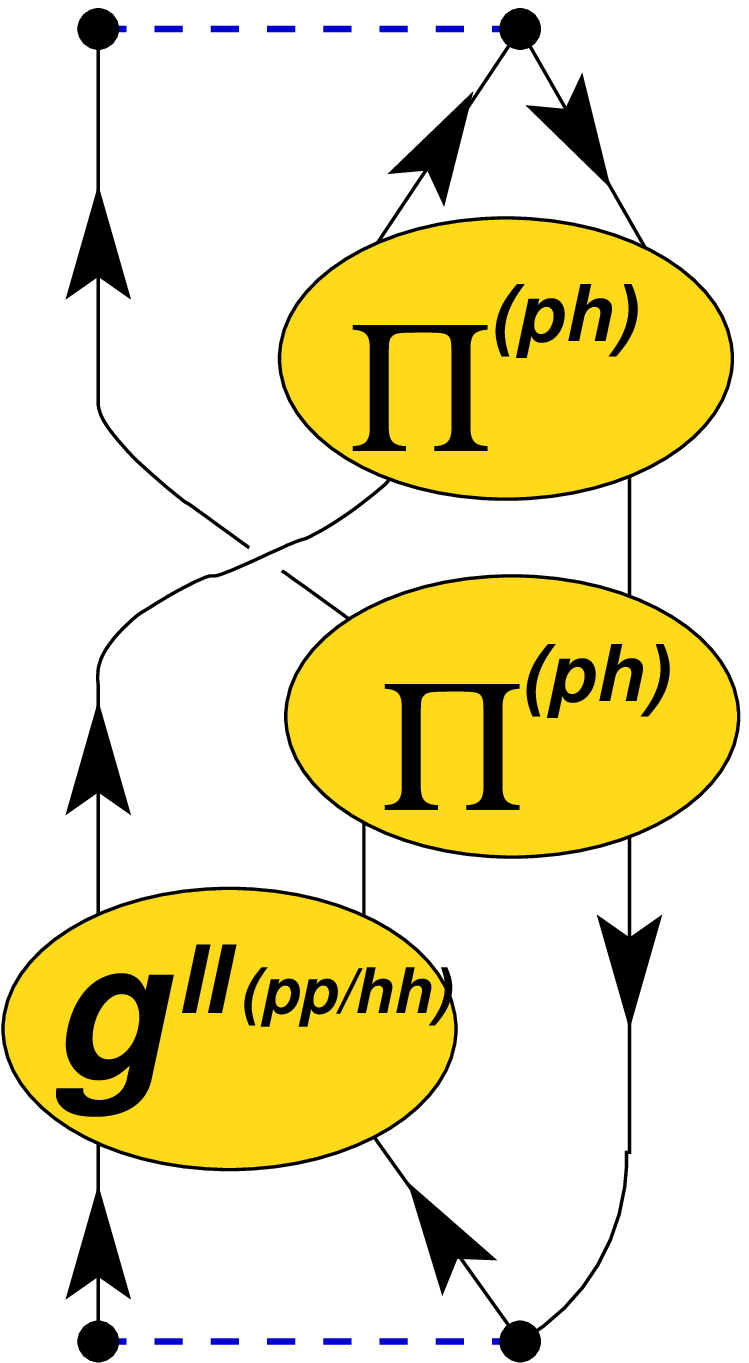,width=.8in} \hspace*{20pt}
            \psfig{file=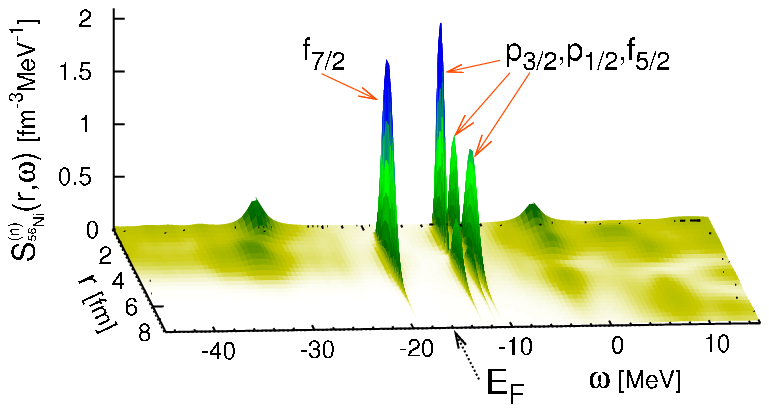,width=3.8in,height=1.5in}}
\caption{\label{fig_frpa}
{\em Left.} One of the self-energy diagrams resummed by the FRPA formalism.
 Arrows up (down) refer to quasiparticle (quasihole) states, the $\Pi^{(ph)}$
propagators include collective \hbox{particle-hole} and charge-exchange resonances, and the $g^{II}$
include pairing between two-particle or two-hole vibrations.
 The FRPA method sums analogous diagrams,
with any numbers of phonons, to all orders.$^{\rm 3,4}$
{\em Right.}  Calculated single-particle spectral function for neutrons in $^{56}$Ni.
 Energies above~(below) E$_F$ are for transitions to excited
states of $^{57}$Ni~($^{55}$Ni). }
\end{figure}

The single-particle spectral function is defined as
\begin{equation}
  S(r,\omega) = 
  \sum_n  |\langle \Psi^{A-1}_n | a_{\bf r} | \Psi^{A}_0 \rangle |^2 
                     \delta(\omega-E^A_0+E_n^{A-1}) \, + \,
  \sum_n  |\langle \Psi^{A+1}_n | a^\dag_{\bf r} | \Psi^{A}_0 \rangle |^2 
                     \delta(\omega-E_n^{A+1}+E^A_0) 
\label{SpFnct}
\end{equation}
which is interpreted as the joint probability of adding or removing a nucleon
at position ${\bf r}$ while leaving the residual system in an eigenstate
$|\Psi^{A\pm1}_n\rangle$ of energy $E_n^{A\pm1}$. Fig.~\ref{fig_frpa} shows
the spectral function for neutrons in $^{56}$Ni,  calculated
in the FRPA scheme.\cite{bar.09c}
The chiral N3LO interaction\cite{Ent.03} was used with a monopole correction
to account for missing three-nucleon forces.
Integrating Eq.~(\ref{SpFnct}) over $\omega\in]-\infty,E_F]$ yields
the matter distribution [the density matrix $\rho({\bf r})$], while
for $\omega>0$ one has elastic scattering states for $n$+$^{56}$Ni.
The QPs associated with the orbits in the $pf$ shell are also visible in the
figure and are normalized to their respective SFs,
\begin{equation}
Z^{A\pm1}_n  = \int d^3 r \; |\langle \Psi^{A\pm1}_n | a_{\bf r} | \Psi^{A}_0 \rangle |^2 =
  \left.
  \frac{1}{  1   -
  \frac{\partial \Sigma^\star_{\hat n \hat n}(\omega)}{\partial \omega}
  } \right\vert_{\omega=\pm(E_n^{A\pm1}+E^A_0)}
\label{eq:sf}
\end{equation}
where $\Sigma^\star_{\hat n \hat n}(\omega)$ is the self-energy calculated for each given QP state,~$n$.

One-nucleon removal cross sections result from a non-trivial folding of Eq.~(\ref{SpFnct}), transition operators, and final state interactions (FSI). Nevertheless, ($e$,$e'p$) reactions in particular kinematics can be dominated by the spectral function\cite{pvbook,bar.04b} and reveal the structure of Fig.~\ref{fig_frpa} in an
unequivocal manner (see, for example, Ref.~\refcite{o16.Sac}).

\begin{table}[h]
\tbl{\label{ta1}
Energies (in MeV) and spectroscopic factors (as a fraction of
    the independent-particle model) for transitions to the {\em pf} valence orbits
    around $^{56}$Ni.$^{\rm 4,10}$
     The fourth and fifth columns correspond to the contributions from SRC only
and to the full FRPA result (including both SRC and LRC).
    The corrections $\Delta Z_\alpha$ are obtained by comparing shell model and FRPA in the $pf$ shell alone.
}
{\begin{tabular}{rlcccccccc}
\hline                        
\hline                        
         &  Quasi-  &  FRPA     &    Exp.\cite{toi.96}   &~&  FRPA     &    FRPA   & $\Delta Z_n$ & FRPA &    \\
         & particle &           &           &~&   (SRC)   &           &                   & +$\Delta Z_\alpha$  &  Exp.\cite{yur.06} \\
         &  orbit   &(10 shells)&            & &(10 sh.) &(10 sh.)&   (pf shell)      &  (10 sh.)      & \\
\hline                        
$^{57}$Ni:
& $\nu 1p_{1/2}$    &  -11.43   &  -9.134   & &   0.96    & 0.63      & -0.02             & 0.61     & \\
& $\nu 0f_{5/2}$    &  -10.80   &  -9.478   & &   0.95    & 0.59      & -0.04             & 0.55     & \\
& $\nu 1p_{3/2}$    &  -12.78   &  -10.247  & &   0.95    & 0.65      & -0.03             & 0.62    &   0.58(11) \\
$^{55}$Ni:
& $\nu 0f_{7/2}$    &  -19.22   &  -16.641  & &   0.95    & 0.72      & -0.03             & 0.69     &  \\
\\
$^{57}$Cu:
& $\pi 1p_{1/2}$    &   -1.28   &  +0.417   & &   0.96    & 0.66      & -0.04             & 0.62     & \\
& $\pi 0f_{5/2}$    &   -0.58   &           & &   0.96    & 0.60      & -0.02             & 0.58     &  \\
& $\pi 1p_{3/2}$    &   -2.54   &  -0.695   & &   0.96    & 0.67      & -0.02             & 0.65     &  \\
$^{55}$Co:
& $\pi 0f_{7/2}$    &   -9.08   &  -7.165   & &   0.95    & 0.73      & -0.02             & 0.71     &  \\
\hline                        
\hline                        
\end{tabular} }
\end{table}

The FRPA formalism allows to separate the contributions of short- and long-range correlations (SRC and LRC) to Eq.~(\ref{eq:sf})\cite{bar.09d}. The SFs obtained 
are given in the fourth and fifth columns of Tab.~\ref{ta1}: LRC are responsible for most of the quenching.
In order to investigate the importance of configuration mixing near the Fermi surface (not included in the FRPA formalism), both SM and FRPA were also calculated in the sole {\em pf} model space and with the same 
modified N3LO interaction.
Tab.~\ref{ta1} shows that the corrections $\Delta Z_n$, due to extra correlations in the SM, are almost negligible in this case. The total results after adding this correction are given by the seventh column and  nicely agree with the experiment.
Similar conclusions were obtained for $^{48}$Ca.\cite{bar.09d}
Thus, the overall quenching of absolute SFs of valence orbits is mainly
explained by the coupling of particles to collective modes,  and requires model spaces
that cannot be approached by standard SM calculations.
It remains clear that SM 
effects have an important impact on open-shell nuclei where they determine
relative SFs (i.e., the fragmentation pattern at low energy).\cite{Tsa.09}

\section*{Two-nucleon emission}

The cross section for the electromagnetic emission of two nucleons 
can be fully written in terms of the scattering
state of the two final nucleons, the electromagnetic current operator
and the two-nucleon overlap function (TOF).
The Pavia model for ($e$,$e'pN$) has been developed over the years
to a sophisticated treatment of the operator: both one-body
(acting on only one nucleon of each correlated pair) and two-body
currents are added. The latter include $\pi$-seagull, pion-in-flight, 
and $\Delta$-isobar terms. See Refs.~\refcite{giu.07,bar.04a} and references therein
for details. FSI are treated using optical models for nucleon-core scattering
and the two-nucleon interaction between the emitted particles at first order.

The information regarding correlations in the initial state is contained 
in the TOF,
$\psi_n({\bf r}_1,{\bf r}_2) \equiv
  \langle \Psi^{A-2}_n | a_{{\bf r}_1} a_{{\bf r}_2} | \Psi^{A}_0 \rangle$.
In the FRPA scheme this is computed solving the RPA equations for two-hole
states and it is a partial step in the calculation of the single-particle
spectral functions [see the propagator $g^{II}(\omega)$ in Fig.~\ref{fig_frpa}, left].

\begin{figure}[h]
\centerline{\hspace*{5pt}  \psfig{file=O16_ee1pn_Shh_comp_pbarn.eps,width=2.3in,height=1.8in} 
            \hspace*{18pt} \psfig{file=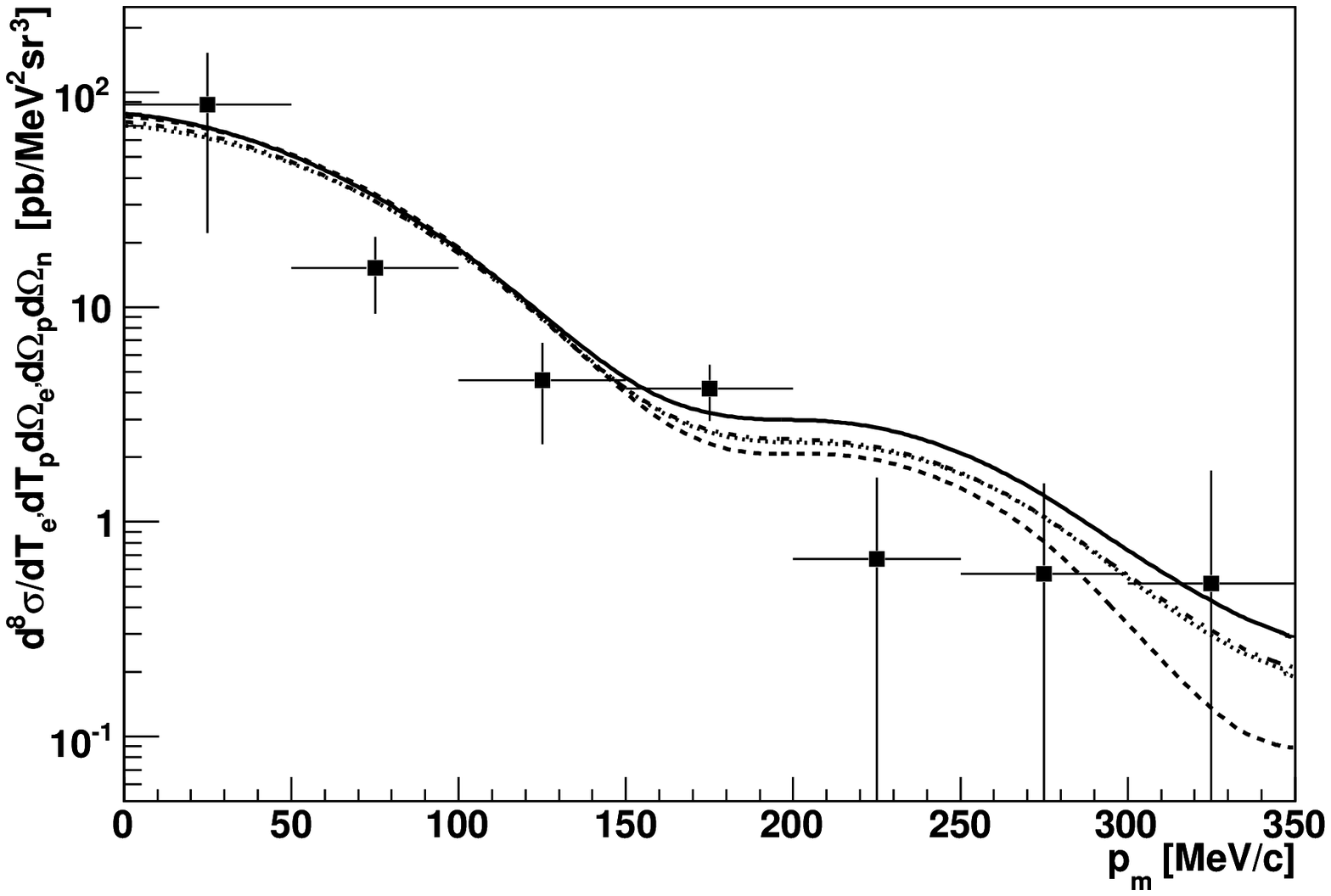,width=2.5in,height=2.in}}
\caption{\label{fig_pn}
{\em Left.} Theoretical cross sections for $^{16}$O($e$,$e'pN$)$^{14}$N
to the $1^+_2$ state as obtained with different models for nuclear
correlations.$^{\rm 14,16,15}$
{\em Right.} FRPA results from Ref.~15, including
all currents (full line), compared to the experiment.$^{\rm 17}$
The data includes both the $1^+_2$, $2^+_1$ and $0^+_1$ final
states of $^{14}$N. Broken lines are obtained by partially neglecting
some two-body currents.$^{\rm 14,17}$ }
\end{figure}

Fig.~\ref{fig_pn} shows the $^{16}$O($e$,$e'pN$) cross section to the 1$^+_2$ excited stated
of $^{14}$N at 3.95~MeV in superparallel kinematics, as obtained with different
approximations for the TOF.
In this particular case LRC effects are dominant. This is seen by comparing to a Jastrow
ansatz for the TOF which includes only central SRC and leads to a clear underestimation
of the cross sections.
 The coupled cluster (including up to double excitations)\cite{giu.99} and
the FRPA\cite{bar.04a} calculations include both LRC and SRC effects. The FRPA 
yields a larger contribution for small missing energies. The latter result is compared
with the experimental cross section of Refs.~\refcite{exp.PN} in Fig.~\ref{fig_pn}.

\vspace*{.5cm}


{\bf Acknowledgments.}
One of the authors (C.B.) thanks M.~Degroote, W.~H.~Dickhoff, T.~Otsuka and D.~Van Neck for several stimulating discussions and for related collaborations.
This work was supported by the Japanese Ministry of Education, Science and Technology (MEXT) under KAKENHI grant no. 21740213.

\end{document}